\documentstyle[12pt,epsfig,fleqn]{article}
  \newcommand{\be}{\begin{eqnarray}}
 \newcommand{\ee}{\end{eqnarray}}
 \newcommand{\nee}{\nonumber\end{eqnarray}}
 \newcommand{\nn}{\nonumber\\}

\def\a{\alpha}

\begin{document}
\begin{center}
{\bf \large Towards a model independent approach to fragmentation
functions}
\end{center}
\begin{center}
{\bf Ekaterina Christova$^1$ and Elliot Leader$^2$}\\
\vspace{0.3cm}
 $^1$ {\it Institute for Nuclear Research and Nuclear Energy, Sofia}\\
  $^2$ {\it Imperial College, London}
\end{center}

We show that the difference cross sections in unpolarized SIDIS $e+N\to e+h+X$ and $pp$ hadron production
$p+p\to h+X$ determine independently in a model independent way, in any order in QCD, the two FFs:
$D_u^{h-\bar h}$ and $D_d^{h-\bar h}$, $h=\pi^\pm , K^\pm$ or a sum over charged hadrons.
If both $K^\pm$ and $K_s^0$ are measured, then  $e^+e^-\to
K+X$, $e+N\to e+K+X$ and $p+p\to K+X$ present independent measurements of just one FF:
$D_{u-d}^{K^++K^-}$. The above results allow to test the existing parametrizations,
obtained with various different assumptions about the FFs, and to
test the $Q^2$ evolution and factorization.

\vspace{.6cm}

\section{Introduction}

There is at present great interest in learning how the spin of the nucleon is built up from the
angular momentum of its constituents. A key ingredient in this is a knowledge of
the polarized parton densities. Most of our knowledge of the polarized PDFs comes from inclusive
deep inelastic scattering (DIS), where, however, one obtains information
only on the combinations $\Delta q(x) + \Delta \bar{q}(x) $. Information on the polarized sea quark
densities can, in principle, be obtained from semi-inclusive deep inelastic scattering (SIDIS)
reactions of the type $l + N \rightarrow l + h + X$, and from semi-inclusive hadron-hadron reactions like
$p + p \rightarrow h + X $. However both of the latter require a knowledge of parton
fragmentation functions (FFs) describing the transition $parton \rightarrow h + X $.

These are not very well known, being obtained principally from analyses of
 $e^+ + e^- \rightarrow h^{\pm } + X $, where, however, only the combinations
 $D_q^h + D_{\bar{q}}^h = D_q^h + D_q^{\bar{h}} \equiv D_q^{h + \bar{h}} $ occur.
 Several sets of FFs are available in the literature -- Kretzer~\cite{Kretzer},
 KKP~\cite{KKP}, AKK~\cite{AKK}), HKNS~\cite{HKNS} etc. One study in~\cite{KrLCh}
 combined $e^+ \, e^-$ data with unpolarized SIDIS data on $ \pi^{\pm } $ production,
 in ~\cite{AKK2008} a combined analysis of $e^+ \, e^-$ and $pp(\bar p)$ data was carried on,
 and quite recently, for the first time, fragmentation functions were extracted
 from a global fit to $e^+ \,e^-, \, \textrm{SIDIS and}  \,
 pp\,\rightarrow \pi ^{\pm} X$  data ~\cite{DSS}. A comprehensive review on the current status of the fragmentation functions
 is presented in~\cite{Albino}.

 Two points should be noted: 1) in all of these analyses (except~\cite{KrLCh})  it was necessary to impose some relations,
 based on theoretical prejudice, between different FFs\footnote{Though in AKK almost no relations were imposed,
   we doubt that data is enough to determine {\it all} FFs independently.}, and  2) there is significant disagreement
 between the various analyses for some FFs.

 For these reasons it is important to try to find ways of extracting FFs without
 any theoretical assumptions about relations between them. In this paper we show
 how information on certain combinations of FFs can be obtained in a model independent
 way from both unpolarized SIDIS and semi-inclusive $pp$ reactions. The key experimental
  ingredients are the differences between cross-sections for producing  hadrons and producing their
  antiparticles i.e data on $ d\sigma^{h-\bar{h}} \equiv d\sigma^h - d\sigma^{\bar{h}} $.
  We are informed that precise data on such observables is feasible~\cite{PHD}.

  Our expressions below correspond to an NLO treatment. LO expressions can be obtained by putting
  $\alpha_s = 0 $, replacing convolutions by ordinary products and using only LO formulae for the partonic cross-sections.

\section{The  cross-sections differences}

 In this section we consider the  cross-sections differences for  the two semi-inclusive processes with charged hadrons $h^\pm$:
\be
e+N \to e+h^\pm +X\quad {\rm and}\quad
p+p\to h^\pm +X
\ee
and define
 \be
\sigma_N^{h^+-h^-} \equiv \sigma_N^{h^+}-\sigma_N^{h^-},\quad
N=p,d\quad {\rm and}\quad \sigma_{pp}^{h^+-h^-} \equiv
\sigma_{pp}^{h^+}-\sigma_{pp}^{h^-}.
\ee
 Using C-invariance of the
strong interactions
 \be
 D_g^{h^+-h^-}=0,\qquad
D_q^{h^+-h^-}=-D_q^{h^+-h^-}\label{C-inv}
\ee we obtain rather
simple expressions for the  cross-section differences and show
that in any order of QCD  they
 are expressed in terms of only non-singlet (NS) combinations of the FFs.

\vspace{.2cm}

\subsection{Unpolarized SIDIS}

 For unpolarized   $e+N \rightarrow e+h^\pm +X$ we obtain
 \footnote{In our formula for the SIDIS cross sections, the common kinematic factors
 have been omitted, see~\cite{wekaons} for the complete expressions.}:
\be
&&\hspace*{-1.6cm} d\sigma_p^{h^+-h^-}(x,z,Q^2) =
\frac{1}{9}\left[ 4u_V\otimes D_u+d_V\otimes  D_d+s_V\otimes D_s\right]^{h^+-h^-}
 \otimes (1+\frac{\a_s}{2\pi} C_{qq})\label{p}\\
&&\hspace*{-1.6cm}d\sigma_d^{h^+-h^-}(x,z,Q^2) =
\frac{1}{9}\left[(u_V+d_V)\otimes(4D_u+ D_d)+2s_V\otimes D_s \right]^{h^+-h^-}
\otimes (1+\frac{\a_s}{2\pi} C_{qq}).\label{d}\nn
\ee
where $u_V$ and $d_V$ are the usual valence quarks
\be
u_V=u-\bar u,\quad d_V=d-\bar d,\quad \textrm{and we define} \quad s_V=s-\bar s\cdot
\ee
Here $x,z,Q^2$ are the usual deep inelastic kinematic variables: $x=Q^2/2P.q=Q^2/2M\nu,\, z=P.P^h/P.q=E^h/\nu$,
 $E$ and $E^h$ are the Lab energies of the incoming lepton and the final hadron, and the $C_{qq}$ are Wilson coefficients.
Note that: 1) $\sigma_N^{h^+-h^-}$ depends only on NS  PDFs and FFs and  is independent of
 the less well known gluon quantities $g(x)$ and $D_g^h$, and 2) the FFs enter multiplied by $q_V=q-\bar q$ which implies that
  the contributions of $D_u^h$ and $D_d^h$ are enhanced by the large
 valence quark densities, while $D_s^h$ is suppressed by the small factor $(s-\bar s)$. Recently a strong bound on
$(s-\bar s)$ was obtained from neutrino experiments, $\vert s-\bar s \vert \leq 0.025$~\cite{Soffer}. This implies that
the large uncertainties in $D_s^h$ should not affect strongly the results for
$\sigma_N^{h^+-h^-}$, and we expect the contribution of $(s-\bar s)D_s^{h^+-h^-}$ to be within the experimental
error and to be neglegible. Then
  Eqs.~(\ref{p}) and (\ref{d}) provide
two independent measurements  for the two unknown FFs $ D_u^{h^+-h^-}$ and $ D_d^{h^+-h^-}$.
These equations hold for the sum over charged hadrons and for each identified hadron separately. \newline

Further information can be obtained when the final hadrons are identified.

1) If $h=\pi^\pm$ then Eqs.~(\ref{p}) and (\ref{d}) present two independent measurements that determine
 $D_u^{\pi^+-\pi^-}$ and $D_d^{\pi^+-\pi^-}$ in a model independent way. Comparison to the existing parametrizations
 for $D_u^{\pi^\pm}$ and $D_d^{\pi^\pm}$ would be a check of these parametrizations.
 Also, comparing $D_u^{\pi^+-\pi^-}$ and $D_d^{\pi^+-\pi^-}$
  would check the usually made assumption
\be
D_u^{\pi^+-\pi^-} = -D_d^{\pi^+-\pi^-}.\label{SU2}
\ee
Recently in \cite{DSS} it was shown, for the first time, that this relation might be violated up to 10 \%.

If eq. (\ref{SU2}) holds, then  Eqs.~(\ref{p}) and (\ref{d}) look particularly simple:
\be
&&\hspace*{-1.6cm} d\sigma_p^{\pi^+-\pi^-} =
\frac{1}{9}\left[ 4u_V-d_V\right]\otimes (1+\frac{\a_s}{2\pi}C_{qq})\otimes  D_u^{\pi^+-\pi^-}\\
&&\hspace*{-1.6cm}d\sigma_d^{\pi^+-\pi^-} =
\frac{1}{3}\left[u_V+d_V\right]\otimes (1+\frac{\a_s}{2\pi} C_{qq})\otimes D_u ^{\pi^+-\pi^-}.
\ee
Thus, if (\ref{SU2})  holds, it should be possible to express $\sigma_p^{\pi^+-\pi^-}$ and $\sigma_d^{\pi^+-\pi^-}$
 in terms solely of a single quantity, $D_u^{\pi^+-\pi^-}$.

2) If $h=K^\pm$ then Eqs.~(\ref{p}) and (\ref{d}) determine in a model independent way
$D_u^{K^+-K^-}$  and
$D_d^{K^+-K^-}$. Comparing $D_u^{K^+-K^-}$  to the existing parametrizations would check the parametrizations for kaon FFs.
In all parametrizations it is always assumed that
\be
D_d^{K^+-K^-}=0.\label{Dd}
\ee
The above measurement would be a direct test of this assumption.

If (\ref{Dd}) holds,
Eqs.~(\ref{p}) and (\ref{d}) look particularly simple:
\be
&&\hspace*{-1.6cm} d\sigma_p^{K^+-K^-} =
\frac{4}{9}u_V \otimes (1+\frac{\a_s}{2\pi}C_{qq})\,\otimes  D_u^{K^+-K^-}
\\
&&\hspace*{-1.6cm}d\sigma_d^{K^+-K^-} =
\frac{4}{9}\left[u_V+d_V\right]\otimes (1+\frac{\a_s}{2\pi} C_{qq})\otimes D_u ^{K^+-K^-}.
\ee
Thus, if (\ref{Dd}) holds, without any other assumptions,
$\sigma_p^{K^+-K^-}$ and $\sigma_d^{K^+-K^-}$ will be determined by a single FF,  $D_u^{K^+-K^-}$.

 Recently, very precise HERMES data
on charged pion and kaon production on unpolarized SIDIS were presented in~\cite{PHD}.
This would allow to construct the discussed cross-section
differences with enough precision. \\
\vspace{.2cm}

\subsection{Unpolarized semi-inclusive hadron-hadron reactions}

 According to the factorization theorem, the general expression for single inclusive production of a hadron $h$
 with high transverse momentum in proton-proton collisions
\be
p\,(P_A) + p\,(P_B) \to h\,(P^h) +X
\ee
  is given by
 \be
 E^h\frac{d\sigma_{pp}^{h}}{d^3P^h} &=&\sum_{a,b,c}\int dx_a\int dx_b\int dzf_a^A(x_a,\mu_F)f_b^B(x_b,\mu_F)D_c^h(z,\mu_F')
 \times\nn
 &&\times \,d\hat\sigma_{ab}^{cX}(x_aP_A,x_bP_B,P^h/z,\mu_R,\mu_F,\mu_F').\label{ppgeneral}
 \ee
Here the sum is over all contributing partonic channels $a+b\to c+X$ and $d\hat\sigma_{ab}^{cX}$ are the corresponding
partonic cross sections [see Eq.~(\ref{partXsections})] calculable
in perturbative  QCD; $\mu_F$, $\mu_F'$ and $\mu_R$ are the factorization scales
associated with the quark densities, fragmentation functions and
renormalization respectively, which, in the following, we take as equal.

Using C-invariance, Eq. (\ref{C-inv}), without any assumptions
about FFs and PDFs, we obtain the following expression for the
cross-section differences valid in any order in QCD:
\be
\hspace*{-1cm}E^h\frac{d\sigma_{pp}^{h^+-h^-}}{d^3P^h} &=&
\frac{1}{\pi}\int dx_a\,\int dx_b\,\int \frac{dz}{z}\times\nn
&&\times \sum_{q=u,d,s}\left[ L_q(x_b,t,u)q_V(x_a)
+L_q(x_a,u,t)q_V(x_b)\right]D_q^{h^+-h^-}(z) \label{ppdiff}
\ee
where we have neglected contributions from heavy quarks since
their contributions are proportional to $c-\bar{c},\, b-\bar{b},\,
t-\bar{t} $ respectively\footnote{Strictly speaking, in NLO eq.(\ref{ppdiff}) is an exact expression only if the masses of
quarks are negligible, $m_q/\sqrt s \ll 1$. For heavy flavour production
  a charge asymmetry  of order $\a_s^3$ is generated~\cite{Halzen,Kuhn}
 and more partonic processes will contribute.
 However,  these effects  seem  too small to affect our considerations.}

 Here
\be
&&\hspace*{-2cm}L_u(x,t,u)= \tilde
u(x)\,d\hat\Sigma(s,t,u)
+[\tilde d(x)+\tilde s(x)]\, d\hat\sigma_{qq'}^{qX}(s,t,u)+ g(x)\, d\hat\sigma_{qg}^{(q-\bar q)X}(s,t ,u)\\
&&\hspace*{-2cm}L_d(x,t,u)=
\tilde d(x)\,d\hat\Sigma(s,t,u)+[\tilde u(x)+\tilde s(x)]\,
 d\hat\sigma_{qq'}^{qX}(s,t ,u)+ g(x)\, d\hat\sigma_{qg}^{(q-\bar q)X}(s,t ,u)\\
&&\hspace*{-2cm}L_s(x,t,u)=
\tilde s(x)\,d\hat\Sigma(s,t,u)+[\tilde u(x)+\tilde d(x)]\,
d\hat\sigma_{qq'}^{qX}(s,t ,u)+ g(x)\, d\hat\sigma_{qg}^{(q-\bar q)X}(s,t ,u)
\ee
where
\be
d\hat\Sigma\equiv \left[\,d\hat\sigma_{qq}^{qX}(s,t ,u)
+ \frac{1}{2}\,d\hat\sigma_{q\bar q}^{(q-\bar q)X}(s,t ,u)\right]\qquad
\tilde q \equiv q+\bar q
\ee

 The partonic cross-section $d\hat\sigma_{ab}^{cX}$ for the inclusive process $a+b\to c+X$ is a
 function of the corresponding Mandelstam variables:
\be \label{partXsections}
 &&d\hat\sigma_{ab}^{cX}(s, t,u)\equiv \frac{d\hat\sigma}{dt}(ab\to cX),\quad s=(p_a+p_b)^2=(x_aP_A+x_bP_B)^2,\nn
 && t=(p_a-p_c)^2=(x_aP_A-p_c)^2,\quad u = (p_b-p_c)^2=(x_bP_B-p_c)^2,\nn
 &&\quad p_c=P^h/z,
 \ee
where $P^h$ stands for $P^{h^\pm}$  respectively. The $d\hat\sigma_{ab}^{cX}$ are calculated in perturbative QCD.
In LO these are $2\to 2$ QCD scattering processes, i.e.
$X$ stands just for one parton, $s+t+u=0$ and one of the integrations can be done immediately.
In total there are 8 different LO cross sections, expressions for
which can be found in many places, for example~\cite{Elliot}.
In NLO, apart from the virtual one-loop
 corrections to the $2\to 2$ processes,
 also  real $2\to 3$
new processes of order $O(\a_s^3)$ are included. This leads to twenty different  inclusive processes~\cite{NLOcorrs},
\cite{Vogelsang2003}.
In this case  $s$, $t$ and $u$ are independent variables.
 Eq. (\ref{ppdiff}) implies that due to C-invariance  only four inclusive partonic
cross sections contribute to $d\sigma_{pp}^{h^+-h^-}$ in LO and six in NLO. These are:
\be
LO: \quad qq'\to q(q'),\quad qq\to q(q),\quad q\bar q\to q(\bar q),\quad qg\to q(g)\nn
NLO: \quad qq'\to qX,\quad qq\to qX,\quad q\bar q\to qX, \bar qX\quad qg\to qX, \bar qX
\ee
where the final $q$ or $\bar q$ are the fragmenting quarks.

The cross section (\ref{ppdiff}) involves only NS  FFs and, thus,
the most troublesome $D_g^h$ does not contribute. Also, we would like to emphasize that
 the structure of the cross section is just the same
as in SIDIS --- $D_q^{h^+-h^-}$ enters always multiplied by
$(q-\bar q)=q_V$, i.e. in the  combination $q_V\,D_q^{h^+-h^-}$.
This implies again that $D_u^{h^+-h^-}$ and $D_d^{h^+-h^-}$ are
enhanced by the large valence-quark densities, while
$D_s^{h^+-h^-}$ is suppressed by the small quantity $(s-\bar s)$
and  its
 contribution should be negligible.   However, that should be checked by calculating
  $L_q(s,t,u)$, which depends only on known quantities.

Thus, $ep$, $ed$ and $pp$ semi-inclusive  cross-section differences determine the same combinations of  FFs:
$D_u^{h^+-h^-}$, $D_d^{h^+-h^-}$ and $D_s^{h^+-h^-}$.
Using the experimentally well justified  approximation $s=\bar s$,
 $D_s^{h^+-h^-}$ will not contribute and the three
 semi-inclusive difference cross sections of $ep$, $ed$ and $pp$  scattering determine
 independently the two quantities
$D_u^{h^+-h^-}$ and $D_d^{h^+-h^-}$.    If $s-\bar s=0$, eq. (\ref{ppdiff}) reads:
\be
\hspace*{-1cm}E^h\frac{d\sigma_{pp}^{h^+-h^-}}{d^3P^h} &=&
\frac{1}{\pi}\int dx_a\,\int dx_b\,\int \frac{dz}{z}\times\nn
&&\times \left\{\left[ L_u(x_b,t,u)u_V(x_a) +(x_a \leftrightarrow x_b, u \leftrightarrow t)\right]D_u^{h^+-h^-}(z)+\right.\nn
&&\hspace*{.5cm}+\left.\left[ L_d(x_b,t,u)d_V(x_a) +(x_a \leftrightarrow x_b, u \leftrightarrow t)\right]D_d^{h^+-h^-}(z)\right\}.
\ee
As $D_q^{h^+-h^-}$ are non-singlets the gluon FF that introduces, in general,
lot of uncertainties will not appear in the $Q^2$-evolution.

Note that taking $s-\bar s=0$ is not an assumption -- it is an approximation linked to the
precision of the experiment. If the accuracy for the  cross-section differences  justifies doing so,
 the strange quark contribution can be included  and the  cross-section differences will provide information about
$(s-\bar s)\,D_s^{h^+ -h^-}$ as well.  It has been suggested that a relatively big $s-\bar s$
difference could be generated in NNLO perturbative QCD ~\cite{s-bars}.

If $h=\pi^\pm$, eq.(\ref{SU2}) implies that $\sigma_{pp}^{\pi^+-\pi^-}$ is expressed solely in terms of
$D_u^{\pi^+-\pi^-}$:
\be
\hspace*{-1cm}E^h\frac{d\,\sigma_{pp}^{\pi^+-\pi^-}}{d^3P^\pi} &=&
\frac{1}{\pi}\int dx_a\,dx_b\,\frac{dz}{z}\,
\left[L_u(x_b,t,u)u_V(x_a) -L_d(x_b,t,u)d_V(x_a)\right. +\nn
&&+\left.L_u(x_a,u,t)u_V(x_b)-L_d(x_a,u,t)d_V(x_b)\right]D_u^{\pi^+-\pi^-}.
\ee
Thus, not only the SIDIS cross sections $\sigma_p^{\pi^+-\pi^-}$ and  $\sigma_d^{\pi^+-\pi^-}$,
but also the single inclusive proton-proton collisions $\sigma_{pp}^{\pi^+-\pi^-}$ are
 expressed  in terms of the single quantity $D_u^{\pi^+-\pi^-}$ if $D_u^{\pi^+-\pi^-}=-D_d^{\pi^+-\pi^-}$ holds.

If $h=K^\pm$  the difference cross sections will determine only $D_u^{K^+-K^-}$ and $D_d^{K^+-K^-}$,
which would test the assumption  $D_d^{K^+-K^-}=0$. If $D_d^{K^+-K^-}=0$, then $d\sigma_{pp}^{K^+-K^-}$
 will be expressed solely in terms of one fragmentation function, $D_u^{K^+-K^-}$.

 Recently, BRAHMS(RHIC)~\cite{BRAHMS} presented data on $\pi^\pm$ and $K^\pm$
 production, that might allow to form the above differences with reasonable
 accuracy.

\vspace{.6cm}

\section{ $K^\pm$ and $K_s^0$ production}

 If in addition to the charged  $K^\pm$ also  neutral kaons $K_s^0=(K^0+\bar K^0)/\sqrt 2$
   are measured,  no new FFs are introduced into the cross-sections. This is a consequence of SU(2) invariance
   of the strong interactions. We have:
\be
&&D_u^{K^+ + K^--2K_s^0}=-D_d^{K^+ + K^--2K_s^0}={(D_u-D_d)}^{K^+ + K^-},\nn
&&
D_s^{K^+ + K^--2K_s^0}= D_c^{K^+ + K^--2K_s^0}=D_b^{K^+ + K^--2K_s^0}=D_g^{K^+ + K^--2K_s^0}=0.
\label{SU2kaons}
\ee
We shall show that the combination
\be
\sigma ^{K^++K^--2K_s^0}\equiv \sigma ^{K^+}+\sigma ^{K^-}-2\sigma ^{K_s^0}
\ee
 in the three types of semi-inclusive processes
 \be
&&e^++e^-\to K+X,\qquad K=K^\pm, K^0_s\label{e+e-Ks}\\
&&e+N\to e +K+X,\qquad N=p,d, \qquad K=K^\pm, K^0_s,\label{SIDISKs}\\
&&p+p\to K+X,\qquad K=K^\pm, K^0_s\label{ppKs}
 \ee
always measures only one  NS combination of FFs, namely
$(D_u-D_d)^{K^++K^-}$.  This result relies  only on SU(2) invariance for the kaons and
does not involve {\it any} assumptions about PD's or  FFs; it holds in any order in QCD. We shall consider the three processes
separately.
\vspace{.2cm}

 Semi-inclusive kaon production
 in $e^+e^-$ and $eN$ scattering was considered earlier in~\cite{wekaons}. For completeness we qoute the results.

\subsection{$e^+ + e^- \rightarrow K + X $}
 For the  $z$-distribution in $e^+e^-\to(\gamma ,Z)\to K+X$ we have
 \footnote{A misprint in the corresponding formula in~\cite{wekaons} has been corrected here.}:
\be
 d\sigma_{e^+e^-}^{K^++K^--2K_s^0}(z,Q^2)=6\,\sigma_0\, (\hat e_u^2-\hat e_d^2)(1+\frac{\a_s}{2\pi} \, C_q\,\otimes \,)
 \,D_{u-d}^{K^++K^-}(z, Q^2)\label{e+e-K}
 \ee
 Here $ \sigma_0=4\pi\alpha_{em}^2/3\,s$ and
 \be
\hat{e_q}^2(s) =e_q^2 - 2e_q
\,v_e\,v_q\,\Re e \,h_Z + (v_e^2 + a_e^2) \, \left[(v_q)^2
+(a_q)^2\right]\, \vert h_Z\vert ^2,
\ee
 where $ h_Z =
[s/(s-m_Z^2+im_Z\Gamma_Z)]/\sin ^2 2\theta_W$,
  $e_q$ is the
 charge of the quark $q$ in units of the proton charge, and, as
  usual,
 \be
 v_e&=&-1/2 +2 \sin^2\theta_W,\quad a_e=-1/2, \nn
v_q&=&I_3^q-2e_q\sin^2\theta_W,\quad a_q=I_3^q, \quad I_3^u = 1/2,
\quad I_3^d = -1/2.
\ee
$z$ is the fraction of the  momentum of the
fragmenting parton transferred to the hadron $h$: $z=
2(P^h.q)/q^2=E^h/E $, where $E^h$ and $E$ are the CM energies of
the final hadron $h$ and the initial lepton, and $\sqrt s=2E$.
\vspace{.2cm}

\subsection{$eN\rightarrow e + K + X $}

 The cross-sections are given by
 \be
d\sigma_p^{K^++K^--2K_s^0}= \frac{1}{9}[(4\tilde u-\tilde d)\otimes  (1+\frac{\a_s}{2\pi}C_{qq}) +
\frac{\a_s}{2\pi}g\otimes  C_{gq}]\otimes
\,D_{u-d}^{K^++K^-}\label{SIDISpK}\\
d\sigma_d^{K^++K^--2K_s^0}=\frac{1}{3} [(\tilde u+\tilde d)\otimes
(1+\frac{\a_s}{2\pi}C_{qq} ) + 2\frac{\a_s}{2\pi} \,g\otimes
C_{gq}]\otimes \,D_{u-d}^{K^++K^-}\label{SIDISdK} \ee
\vspace{.2cm}

\subsection{$pp\rightarrow K + X $}

  From Eq.~(\ref{ppgeneral}), using (\ref{SU2kaons}), for $pp\to K+X$, $K=K^\pm, K^0_s$ we obtain:
\be
\hspace*{-1cm}E^K\frac{d\sigma_{pp}^{K^++K^--2K_s^0}}{d^3P^K}&&=\frac{1}{\pi}
\sum_{a,b}\int\, dx_a\int\, dx_b\int\, \frac{dz}{z}\,\times\nn
&&\hspace*{-2cm}\times f_a^{A}(x_a)f_b^{B}(x_b)
\left[\,d\sigma_{ab}^{uX}+d\sigma_{ab}^{\bar uX}-
d\sigma_{ab}^{dX}-d\sigma_{ab}^{\bar dX}\right] D_{u-d}^{K^++K^-}(z)\label{ppK}
\ee
Here the sum over $a,b$ is over all partons.

It is remarkable that all three processes measure the same NS $D_{u-d}^{K^++K^-}$.
This implies, in particular, that even if one does not know the combination $D_{u-d}^{K^++K^-}$,
it should be possible to fit the data on all four  processes solely with one NS fragmentation function, whose evolution
 will not introduce any other FFs.

Written in detail Eq.~(\ref{ppK}) reads:
\be
\hspace*{-1cm}E^K\frac{d\sigma_{pp}^{K^++K^--2K_s^0}}{d^3P^K}&&=\frac{1}{\pi}
\int\, dx_a\int\, dx_b\int\, \frac{dz}{z}\,\times\nn
&&\hspace*{-1.5cm}\times\left\{\left[ \tilde u(x_a)[\tilde d(x_b) + \tilde s(x_b)] -
\tilde d(x_a)[\tilde u(x_b) + \tilde s(x_b)]\right]\hat\sigma_{qq'}^{qX}(s,t,u)\right.+\nn
&&\hspace*{-1.5cm}+2\left[u(x_a)u(x_b)+\bar u(x_a)\bar u(x_b) -
[d(x_a)d(x_b)+\bar d(x_a)\bar d(x_b)]\right]\hat\sigma_{qq}^{qX}(s,t,u)+\nn
&&\hspace*{-1.5cm}+\left[d(x_a)\bar d(x_b)-u(x_a)\bar u(x_b)\right]
\left[2\hat\sigma_{q\bar q}^{q'X}(s,t,u)-\hat\sigma_{q\bar q}^{(q+\bar q)X}(s,t,u)\right]+\nn
&&\hspace*{-1.5cm}+[\tilde u(x_a)-\tilde d(x_a)]g(x_b)\left[\hat\sigma_{qg}^{(q+\bar q-2q')X}(s,t,u)\right]\nn
&&\hspace*{-1.5cm}
+\left.\left[(x_a\leftrightarrow x_b),(t\leftrightarrow u)\right]\right\}D_{u-d}^{K^++K^-}(z).
\ee
 Note that only
 8 inclusive processes (5 in LO) contribute. This result is readily obtained using the symmetry properties of the partonic
 cross sections and Eq.~(\ref{SU2kaons}).

 The BRAHMS data on $K^\pm$-production, combined with the data on $K^0_s$-production from STAR(RHIC)~\cite{STAR}
 may allow to form $\sigma_{pp}^{K^++K^--2K_s^0}$ with reasonable accuracy..
\vspace{.3cm}

In this section we have presented four independent measurements, Eqs.~(\ref{e+e-Ks}), (\ref{SIDISKs}) and
(\ref{ppKs}), that determine in a model independent way
 the NS combination of the kaon  FF $D_{u-d}^{K^++K^-}$. As these expressions are  model independent,
 it would be interesting to compare the resulting FF  to the existing parametrizations extracted from
$e^+e^-$ data, which were obtained with various assumptions.
In addition, Eqs.~(\ref{e+e-K})- (\ref{SIDISdK}) allow one to
compare the FF obtained in $e^+e^-$
at rather high $Q^2\simeq m_Z^2$, Eq.~(\ref{e+e-K}),
with those from SIDIS at quite low $Q^2$, Eqs.~(\ref{SIDISpK}) -- (\ref{SIDISdK}). Comparing with an extraction based on
Eq.~(\ref{ppK}) would provide a challenging test of the universality of the FFs.
\vspace{.3cm}

Note that the analogous combination for pions -- $\sigma^{\pi^++\pi^--2\pi^0}$,
 for all three types of processes is  identically zero with the usually used assumptions
 $D_c^{\pi^++\pi^-}=2D_c^{\pi^0}$.


\section{Conclusions}

Three types of experiments involving high energy collisions of elementary particles
with unpolarized beams have been studied: $e^+e^-\to h+X$, SIDIS $eN\to e+h+X$ ( $N=p$  and $d$) and  $pp\to h+X$.
Based only on  factorization
 and C-invariance of the strong interactions, and without any assumptions about PDFs and FFs, we show that
in any order in QCD the difference cross sections $\sigma^{h^+}-\sigma^{h^-}$ for SIDIS and for
 $pp\to h+X$ are expressed in terms of the same non-singlet FFs $D_{u,d,s}^{h^+-h^-}$. If in addition to  charged kaons
 $K^\pm$, also the neutral $K^0_s$ can be measured, then SU(2) invariance  implies
 that for all three types of process the combination $\sigma^{K^++K^--K^0_s}$ is
  expressed solely in terms of one non-singlet combination
  $(D_u-D_d)^{K^++K^-}$. These measurements do not provide
   full information about the FFs, but only part of it, which
  however is model independent and correct in any order in QCD.
  This  allows to test both  the existing parametrizations
  and some of the usually made assumptions. They also provide a test  of $Q^2$-evolution and factorization.


\section*{Acknowledgements}

One of us (E.Ch.) is thankful to Marco Stratmann for the useful discussions.


\end{document}